\documentclass[10pt,twocolumn]{revtex4}
\usepackage{graphicx}
\usepackage{amssymb}
\usepackage{amsmath}
\usepackage{bm}
\usepackage{color}

\begin{document}

\title{Dual backgrounds and their stability during $\bm{\chi^{(2)}}$ comb generation in microresonators}
\author{B.\ Sturman$^1$, E.\ Podivilov$^1$, J.\ Szabados$^2$ and I.\ Breunig$^{2,3}$ }
\affiliation{	
$^1$Institute of Automation and Electrometry, Russian Academy of Sciences, 630090 Novosibirsk, Russia \\
$^2$Laboratory for Optical Systems, Department for Microsystems Engineering–IMTEK, University of Freiburg,
Georges-K\"{o}hler-Allee 102, 79110 Freiburg, Germany \\
$^3$Fraunhofer Institute for Physical Measurement Techniques IPM, Georges-K\"{o}hler-Allee 301, 79110 Freiburg, Germany 
}

\begin{abstract}
Light states relevant to the $\chi^{(2)}$ comb generation in optical microresonators possess typically dual backgrounds for coupled first-harmonic (FH) and second-harmonic (SH) envelopes. Stability and/or instability of these backgrounds is crucial for realization of stable $\chi^{(2)}$ combs and also for an efficient SH generation. We explore the properties of the dual backgrounds and their instability for the cases of FH and SH pumping of the resonator. In contrast to the optical parameteric oscillation, the instability is controlled by a 4th-degree characteristic equation for the increment. Coefficients of this equation depend not only on wavenumbers of the perturbations, the pump power, and dispersion parameters, but also on FH-SH group velocity difference (temporal walk-off). Our results include characterization of the regions and conditions of stability for the FH- and SH-pumping cases and different spectral ranges. 
\end{abstract}

\maketitle

\section{Introduction}\label{S1}

Nonlinear optics of high-$Q$ optical microresonators is nowadays a prosperous research area with numerous applications~\cite{VahalaNature03,Matsko03,Review16,IngoReview16}. Generation of frequency combs and temporal solitons have become especially important during the last decade~\cite{KippNP14,Vahala15,KippScience18}. Until recently, the main efforts with this subject were focused on $\chi^{(3)}$ resonators~\cite{HerrNP17,KerrReview19,ChangNC20}. The key notion gained is that the comb and soliton generation are interrelated and mutually reinforcing~\cite{KippScience18}. Transferring the comb-soliton concept to $\chi^{(2)}$ resonators is nowadays in demand. Numerous theoretical and experimental results were obtained here during the last years~\cite{Att1,Att2,Att3,WabnitzPRA16,Att4,Wabnitz18,Skryabin19,Skryabin19A,WePRA20,WeOE20,WeOE21,WeJOSAB20,IngoPRL20,IngoAPL20,HendryOL20,NP21}. In theory, they are relevant to exploration (usually numerical) of different comb and soliton regimes. Substantial differences between microresonators and bulk singly resonant cavities were found. No experimental evidences of net $\chi^{(2)}$ comb-soliton generation were obtained so far. 

Complexity of the subject compared to the $\chi^{(3)}$ case is caused by the simultaneous involvement (interference) of two basic nonlinear processes -- the thresholdless second harmonic (SH) excitation and the threshold-possessing optical parametric oscillation (OPO)~\cite{Matsko03,Review16,IngoReview16}. As a consequence of the presence of coupled FH and SH light envelopes, we have dual comb-soliton characteristics. The latter are controlled by two dispersion coefficients and also by the FH-SH group velocity difference (the temporal walk-off). As the resonator is pumped, any steady-state regime involves a  balance between the modal gain and losses. Also, FH-SH phase matching has to be ensured~\cite{RadialPoling1,RadialPoling2,NaturalPM}. Of special interest is the phase-matching adjustment to the zero walk-off point~\cite{WeOE20,WeOE21}.

\begin{figure}[h]
	\centering    
	\includegraphics[width=7cm]{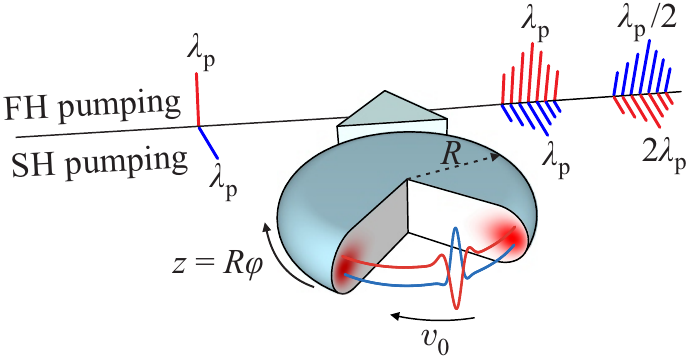}
	\caption{FH and SH pumping schemes for the comb generation; $R$ and $\varphi$ are the major radius and the azimuth angle. The output comb spectra are due to the FH-SH solitons propagating along the resonator rim with a common velocity $v_0$ and possessing intensity backgrounds. Red spots indicate localization of the resonator modes at the rim. The shown polarization scheme is illustrative. }\label{Geo}
\end{figure}

Two main schemes of selective monochromatic pumping, namely pumping into a FH or SH resonator mode, are common, see Fig.~\ref{Geo}. Despite an apparent similarity, they lead to strongly different consequences for the comb-soliton states. In particular, the predicted antiperiodic states, possessing remarkable robustness and accessibility~\cite{WePRA20,WeOE20,WeOE21}, are possible only for the SH pumping. 

Despite the complexity and diversity of $\chi^{(2)}$ comb-soliton phenomena, one of their features, providing deep insights into the subject, is relatively simple. It is the presence of dual backgrounds, i.e. of spatially uniform FH-SH states involving two resonator modes. The properties of these backgrounds are different for the FH and SH pumping schemes. In the FH pumping case, the dual background arises inevitably already at small pump powers. In the SH pumping case, the dual background arises, like comb-soliton states, above a power threshold owing to the OPO. 

\vspace*{-4mm}

Stability and instability of the dual backgrounds has little in common with the conventional OPO. While OPO processes involve (in addition to the pump) only two signal and idler modes~\cite{Matsko03,Review16,IngoReview16}, the background instability deals with four weak FH-SH amplitudes and involves a substantially different set of resonator parameters. The role and properties of this modulation instability are also different for the FH and SH pumping cases. In the FH case, the instability is necessary to initiate the soliton-comb formation. On the other hand, it is harmful in the studies of efficient SH generation~\cite{Review16,IngoReview16,IngoAPL20,SHG-Jan21,SHG-21}. In the SH pumping case, stability of the dual background is closely related to stability of the soliton-comb generation. 
Non-OPO nature of the dual background instability is thus not well recognized, see~\cite{Ikuta20,IngoAPL20,HendryOL20}.   

The purpose of this work is to characterize the dual backgrounds for the cases of FH and SH pumping and investigate their modulation instability. In the first case, we are mostly interested in the determination of the minimum instability thresholds versus the variable experimental parameters and comparison with the OPO properties. In the second case we are interested also in stability properties at high pump powers. While investigations of the modulation instability in high-$Q$ microresonators have some prerequisites, see, e.g.~\cite{WabnitzPRA16,Skryabin19,SkryabinReview}, this subject remains almost unexplored. Our theory is within the mean field approach which is conventional for microresonators. This approach can also be validated by comparison with more general results of~\cite{JOSA11,JOSA12}.  

\section{Initial equations}\label{S2}

The initial nonlinear equations for the FH and SH envelopes, $F = F(z,t)$ and 
$S = S(z,t)$, read~\cite{WePRA20,WeOE20}:
\begin{eqnarray}\label{Initial}
\hspace*{-3mm} ({\rm i} \partial_t + {\rm i}v_1 \partial_z + d_1\partial^2_z - \Omega_1) F &=& 2\mu SF^* + {\rm i}h_1  \nonumber \\
\hspace*{-3mm}({\rm i}\partial_t  + {\rm i}v_2 \partial_z + d_2\partial^2_z  - \Omega_2) S &=& \mu F^2 + {\rm i}h_2 \; . 
\end{eqnarray}
Here $z = R\varphi$ is the rim coordinate, $R$ is the major radius, $\varphi$ is the azimuth angle, $\mu$ and $h_{1,2}$ are positive nonlinear coupling and pumping coefficients, $v_{1,2}$ are the FH and SH group velocities, $d_{1,2}$ are the dispersion parameters of dimension cm$^2$/s~\cite{FootNote}, $\Omega_{1,2} = \Delta_{1,2} - {\rm i}\gamma_{1,2}$, $\Delta_{1,2}$ are the frequency detunings (see also below), and $\gamma_{1,2}$ are the modal decay rates. We suppose that either $h_2$ or $h_1$ is zero; this means that either a FH or SH mode is pumped. The squared values $h_{1,2}^2$ are proportional to the pump power. The modal quality factors are $Q_{1,2} = \pi c/\gamma_{1,2}\lambda_{1,2}$, where $c$ is the speed of light and $\lambda_{1,2}$ are the FH and SH vacuum wavelengths ($\lambda_1 = 2\lambda_2$). All introduced parameters are controllable in experiment. The initial equations incorporate gain and losses, they are dissipative. 
\begin{figure}[h]
	\centering    
	\includegraphics[width=8.4cm]{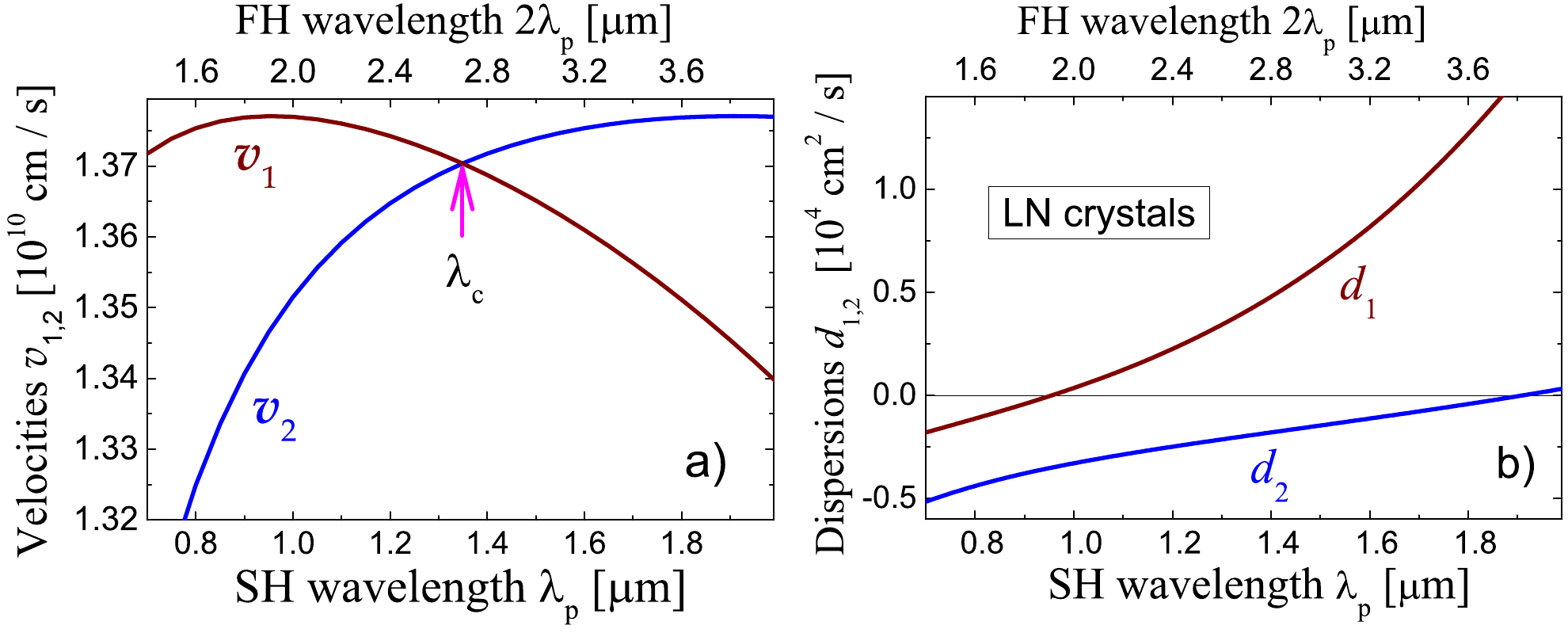}
	\caption{Wavelength dependences $v_{1,2}(\lambda_p)$ (a) and $d_{1,2}(\lambda_p)$ (b) for
		LiNbO$_3$ crystals and the extraordinary polarization of the modes. The arrow indicates zero walk-off point $\lambda_c \simeq 1349$~nm.}\label{vq}
\end{figure}
Figure~\ref{vq} shows the wavelength dependences of $v_{1,2}$ and $d_{1,2}$ for lithium niobate crystals and the extraordinary light polarization relevant to the strongest nonlinear coupling. One sees that parameters $v_{12} \equiv v_1 - v_2$ and $d_{1,2}$ are sign-changing. Eqs.~(\ref{Initial}) and Fig.~\ref{vq} imply the FH-SH phase matching or quasi-phase matching~\cite{Review16,IngoReview16}. To adjust the pump wavelength $\lambda_p$ to an arbitrary spectral point, proper radial poling of the resonator is required~\cite{RadialPoling1,RadialPoling2,WePRA20}. Otherwise, we are restricted to the point of natural (birefringent) phase matching~\cite{NaturalPM} relevant to interaction of the ordinarily and extraordinarily polarized modes and strong walk-off effects. 

Set~(\ref{Initial}) is written for a static coordinate frame. It is useful to rewrite it for a frame moving with the FH velocity $v_1$. To do so, it is sufficient to drop the term with $\partial_z$ in the first equation and replace $v_2$ by $-v_{12}$ in the second one. The velocity difference $v_{12}$ is then the only parameter accounting for the temporal walk-off.

Some comments about the status and properties of the initial equations are useful. \\
-- It is implied that only the azimuth modal numbers matter. This simplifying assumption is common for comb modeling, and it is supported by the results obtained for $\chi^{(3)}$ resonators~\cite{HerrNP17,KerrReview19,ChangNC20}. \\
-- For FH pumping ($h_1 \neq 0$, $h_2 = 0$) we have $\Delta_1 = \Delta_p = \omega_1 - \omega_p$ and $\Delta_2 = 2\Delta_p + \Delta_0 = \omega_2 - 2\omega_p$, where the mismatch $\Delta_0 = \omega_2 - 2\omega_1$ characterizes slightly imperfect phase matching, $\omega_{1,2}$ are the WGM frequencies relevant to the FH and
SH, and $\omega_p$ is the pump frequency.  \\
--  For SH pumping ($h_1 = 0$, $h_2 \neq 0$) we have $\Delta_2 = \Delta_p = \omega_2 - \omega_p$ and $\Delta_1 = (\Delta_p - \Delta_0)/2 = \omega_1 - 0.5\omega_p$. \\
-- Detunings $\Delta_{1,2}$ have to be much smaller in value than the free spectral range $c/nR \sim 10^{11}$~s$^{-1}$, where $n$ is the relevant refractive index. With this restriction they can be regarded as important variable experimental parameters to access different stable background states and also comb-soliton states~\cite{Wabnitz18,Skryabin19,Skryabin19A,WeOE21}. 

\section{Background solutions}\label{S3}

Set~(\ref{Initial}) admits spatially uniform steady-state solutions $\bar{F},\bar{S}$. These background states are  not influenced by $v_{1,2}$ and $d_{1,2}$ and different for the FH and SH pumping cases. We consider them separately.

\begin{figure}[h]
	\centering    
	\includegraphics[width=8.6cm]{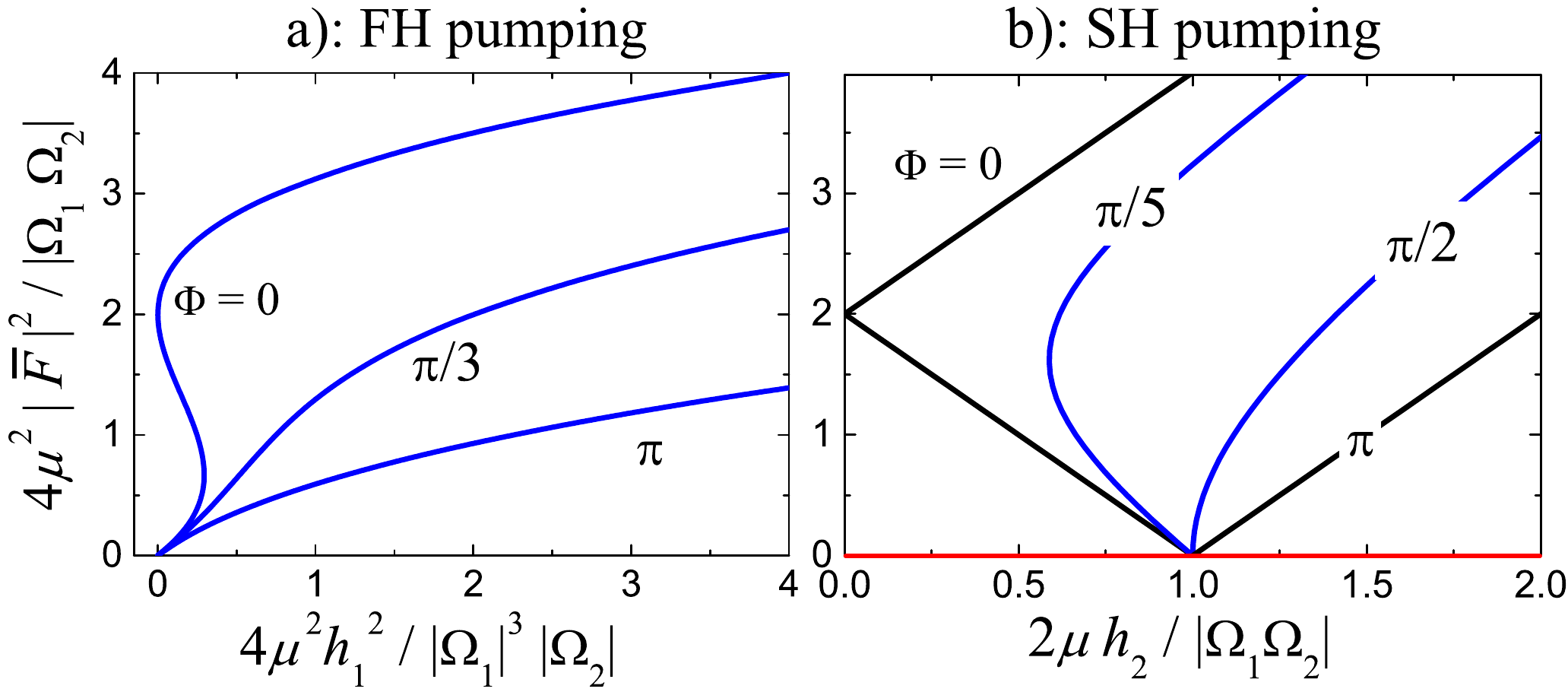}
	\caption{Normalized dependences of the FH background intensity $|\bar{F}|^2$ on the pump strength parameter $h$ for the cases of FH (a) and SH (b) pumping. The curves are relevant to different values of the phase $\Phi = \arg(\Omega_1\Omega_2)$. Zero detuning case, $\Delta_{1,2} = 0$, corresponds to $\Phi = \pi$. The horizontal red line in b) indicates the single background state, $\bar{F} = 0$, $\bar{S} = -ih_2/\Omega_2$.}\label{Backgrounds-FH-SH}
\end{figure}

{\bf FH pumping}: Here we set $h_2 = 0$ to get 
\begin{equation}\label{F_01}
\bar{S} = -\mu\bar{F}^2/\Omega_2, \quad \left(2\mu^2|\bar{F}|^2/\Omega_2 - \Omega_1 \right) \bar{F} = {\rm i}h_1 \,.
\end{equation}
The dependence of $\mu^2h_1^2$ on $x = 2\mu^2 |\bar{F}|^2/|\Omega_1\Omega_2|)$ is cubic,
\begin{equation}\label{BackgroundCubic}
\big[(x - \cos\Phi)^2 + \sin^2\Phi \big] x = 2\mu^2h_1^2/|\Omega_1|^3|\Omega_2| \;.
\end{equation}
It is controlled by the phase $\Phi = \arg(\Omega_1\Omega_2)$~and $|\Omega_{1,2}|$.~The corresponding normalized dependences $|\bar{F}|^2(h_1^2,\Phi)$ are presented in Fig.~\ref{Backgrounds-FH-SH}a. Zero-detuning case, where $\Omega_{1,2} = -{\rm i}\gamma_{1,2}$, corresponds to $\Phi = \pi$ and the slowest single-valued dependence. For $|\Phi| < \pi/6$ the dependence $|\bar{F}|^2(h_1^2)$ becomes three-valued. The phase $\Phi = 0$ can be achieved only asymptotically for $|\Delta_{1,2}|/\gamma_{1,2} \to \infty$.

{\bf SH pumping}: Here we set $h_1 = 0$. Obviously, we have a single background solution $\bar{F} = 0$, $\bar{S} = -{\rm i}h_2/\Omega_2$. It is stable against small perturbations below the OPO threshold, $2\mu h_2 < |\Omega_1\Omega_2|$, see also below. Additionally, there are two branches with $\bar{F} \neq 0$ relevant to the dual background. For them we have
\begin{equation}\label{F_02}
\hspace*{-1mm}\bar{S} \hspace*{-0.5mm} = \hspace*{-0.5mm} -\Omega_1\bar{F}/2\mu\bar{F}^*, \hspace*{1.5mm} \left( 2\mu^2|\bar{F}|^2 \hspace*{-0.5mm} - \hspace*{-0.5mm} \Omega_1\Omega_2 \right)\bar{F} \hspace*{-0.5mm} = \hspace*{-0.5mm} -2{\rm i}\mu h_2\bar{F}^* .
\end{equation}
Each branch is defined here up to the sign: $\bar{F}$ and $-\bar{F}$ are two equivalent solutions relevant to the same $\bar{S}$. Taking the absolute values squared in Eqs.~(\ref{F_02}), we obtain that $|\bar{S}|^2 = |\Omega_1|^2/4\mu^2$ and
\begin{equation}\label{F_0^2}
2\mu^2|\bar{F}|^2_{\pm} = |\Omega_1\Omega_2|\cos \Phi \pm \sqrt{4\mu^2h_2^2 - |\Omega_1\Omega_2|^2\sin^2\Phi } \,;
\end{equation}
as earlier, $\Phi = \arg(\Omega_1\Omega_2)$. Eq.~(\ref{F_0^2}) is supplemented by the obvious conditions $4\mu^2 h_2^2 > |\Omega_1\Omega_2|^2\sin^2\Phi$ and $|\bar{F}|_{\pm}^2 > 0$. In particular, the branch $|\bar{F}|^2_-$ is not always present, and it can exist only within a window of $h_2$. The normalized dependences $|\bar{F}|^2(h_2,\Phi)$ are illustrated by Fig.~\ref{Backgrounds-FH-SH}b. At $\Delta_{1,2} = 0$, when $\Phi = \pi$, there is only one branch $\mu^2|\bar{F}|^2_+ = \mu h_2 - \gamma_1\gamma_2/2$. For $|\Phi| < \pi/2$, the dependence $|\bar{F}|^2(h_2)$ becomes two-valued. 

Note that our representation of the dual FH backgrounds in Fig.~\ref{Backgrounds-FH-SH} employs normalizations making the intensity dependences two-parametric. Without these normalizations, the intensity dependences on the variable parameters look more complicated, see also Sect.~\ref{S4}.  

In the case of multivalued dependence $|\bar{F}|^2(h_{1,2})$, slow increase and decrease of the pumping coefficients $h_{1,2}$ (of the pump power) can be accompanied by hysteresis phenomena. Multivalued sections of curves presented in Fig.~\ref{Backgrounds-FH-SH} can also be accessed via slow changes of the frequency detunings $\Delta_{1,2}$. These sections have to be temporally stable, see below, otherwise they are not physical. 

The difference between the FH and SH pumping cases is not only in particularities of the above graphs and expressions for $\bar{F}$ and $\bar{S}$. In the FH pumping case the background is always dual, i.e. both $\bar{F}$ and $\bar{S}$ are nonzero. In the SH case, in addition to the dual background, there is a single background $\bar{F} = 0$, $\bar{S} \neq 0$. 

All background solutions belong indeed to the class of the periodic states -- antiperiodic background solutions do not exist. Nonetheless, all known strongly localized antiperiodic soliton states show a close relation to the background states~\cite{WeOE20,WeOE21}: Each antiperiodic soliton solution $F(z),S(z)$ tends asymptotically, when increasing the left and right distances from the soliton core, to $\bar{S},\pm\bar{F}$. The corresponding asymptotic values of $|F(z)|^2$ and $|S(z)|^2$ are just the background values $|\bar{F}|^2$ and $|\bar{S}|^2$. 

\section{Main characteristic equation}\label{S4}

Now we represent the FH and SH amplitudes as $F = \bar{F} + \delta F$ and $S = \bar{S} + \delta S$, where $\delta F(z,t)$ and $\delta S(z,t)$ are small perturbations. Owing to $2\pi R$ periodicity, the latter can be expanded in Fourier series,
\begin{equation}\label{Fourier}
\delta F = \sum\limits_{k} F_k \exp({\rm i}kz), \quad \delta S = \sum\limits_{k} S_k \exp({\rm i}kz) 
\end{equation}
with discrete wavenumbers $k = j/R$, where $j = 0,\pm 1, \ldots$ is the mode number. Substituting these modal expansions into Eqs.~(\ref{Initial}), we obtain that four independent quantities, $F_k$, $F^*_{-k}$, $S_k$, $S^*_{-k}$, are mutually coupled within the linear approximation. The corresponding set of dynamic equations is
\begin{eqnarray}\label{FourHarmonics}
{\rm i}\dot{F}_k \hspace*{-1.6mm} &-& \hspace*{-1.6mm}(\Omega_1 + d_1k^2)F_k = 2\mu(\bar{S}F_{-k}^* + \bar{F}^*S_k) \nonumber  \\
-{\rm i}\dot{F}_{-k}^* \hspace*{-1.6mm} &-& \hspace*{-1.6mm} (\Omega^*_1 + d_1k^2)F^*_{-k} = 2\mu(\bar{S}^*F_{k} + \bar{F}S^*_{-k}) \nonumber \\
{\rm i}\dot{S}_k \hspace*{-1.6mm} &-& \hspace*{-1.6mm} (d_2k^2 - v_{12}k + \Omega_2)S_k = 2\mu \bar{F} F_k  \\
-{\rm i}\dot{S}^*_{-k} \hspace*{-1.6mm} &-& \hspace*{-1.6mm} (d_2k^2 + v_{12}k + \Omega^*_2)S^*_{-k} = 2\mu \bar{F}^*F^*_{-k}, \nonumber
\end{eqnarray}
where dots indicate differentiation in $t$. Remarkably, this set does not include $h_{1,2}$. This means that we treat so far uniformly the FH and SH pumping cases. The presence of two terms in the right-hand sides of the first two equations says about interference of processes relevant to the sum and difference frequency generation. 

Setting now $F_k,F^*_{-k},S_k,S^*_{-k} \propto \exp(-{\rm i}\nu t)$, we obtain four linear algebraic equations for the determination of the increment $\nu$. Combining them, it is not difficult to come to a single characteristic equation for $\nu$ in the form
\begin{equation}\label{DispersionEquation}
\hspace*{-1mm} (L^+_1L^+_2 - 4\mu^2|\bar{F}|^2)(L^-_1L_2^- - 4\mu^2|\bar{F}|^2) \hspace*{-0.5mm} = \hspace*{-0.5mm} 4\mu^2|\bar{S}|^2L^+_2L^-_2 .
\end{equation}
Parameters $L_{1,2}^{\pm}$ relevant to a coordinate frame moving with the FH group velocity $v_1$ are 
\begin{eqnarray}\label{L1,2}
L_1^{\pm} &=& \pm (\nu + {\rm i}\gamma_1) - d_1k^2 - \Delta_1 \\
L_2^{\pm} &=& \pm (\nu + {\rm i}\gamma_2 + v_{12} k) - d_2k^2 - \Delta_2 \,.   \nonumber
\end{eqnarray}
Only squared modules $|\bar{F}|^2$ and $|\bar{S}|^2$ enter Eq.~(\ref{DispersionEquation}). They are not independent and expressible by $\mu h_1$ (or $\mu h_2$) and detunings $\Delta_{1,2}$. The form of the corresponding expressions depends on whether the first or second harmonic is pumped and also on which of the background branches is under study. 

Characteristic equation~(\ref{DispersionEquation}) is generally complex, of 4th power in $\nu$ and of 8th power in $k$ (or, equivalently, $j$). It is drastically different from a quadratic characteristic equation relevant to the OPO case~\cite{,Matsko03,Review16,IngoReview16}. The instability occurs for $\nu'' = {\rm Im} (\nu) > 0$, the threshold equation is thus $\nu'' = 0$.  Each solution for $\nu$ can be considered as a complex function of $j$. Using Eqs.~(\ref{DispersionEquation}) and (\ref{L1,2}), one can find that $\nu(j) = -\nu^*(-j)$. The imaginary part of the increment, responsible for the instability, is thus even in~$j$, $\nu''(j) = \nu''(|j|)$, while $\nu'(j) = -\nu'(-j)$. 

Numerical determination of $\nu$ for each pumping scheme and for each particular set of the variable parameters [$h_{1,2}$, $v_{12}(\lambda_p)$, $d_{1,2}(\lambda_p)$, $\gamma_{1,2}$, $\Delta_{1,2}$, and $j = kR$] is not difficult. However, mapping of the instability properties (including the minimum in $j$ thresholds $h_{1,2}^{\rm th}$) in the space of the variable parameters and determination of the main functional dependences are far from trivial. 

Set~(\ref{FourHarmonics}) can also be used for numerical modeling of the temporal evolution of $F_k, F^*_{-k}, S_k, S^*_{-k}$ with arbitrary small initial values. In this way, the values of $\nu''$ can be determined independently for any input parameters to verify the results of our analysis of Eq.~(\ref{DispersionEquation}). 

Substantial simplifications of Eq.~(\ref{DispersionEquation}) are possible. A quite general and not very obligative simplification is the following: According to Eqs.~(\ref{L1,2}), the increment $\nu$ enters the characteristic equation in the combinations $\nu + i\gamma_{1,2}$. In the case of equal decay constants, $\gamma_2= \gamma_1$, we can get rid of $\gamma_{1,2}$ by changing from $\nu$ to $\nu + {\rm i}\gamma_1$. The characteristic equation becomes then real. For other simplifications, we consider different particular and limiting cases.

{\bf Normalization issues:} Proper normalization is practical for the subsequent considerations. Below we apply the following normalized quantities: the normalized background intensities $I_1 = 4\mu^2 |\bar{F}|^2/\gamma_1\gamma_2$ and $I_2 = 4\mu^2 |\bar{S}|^2/\gamma_1^2$, the normalized detunings $\delta_{1,2} = \Delta_{1,2}/\gamma_{1,2}$, the normalized walk-off parameter $\alpha = v_{12}/\gamma_2 R$, the normalized dispersion parameters $\beta_{1,2} = d_{1,2}/\gamma_{1,2} R^2$, and the decay rate ratio $r = \gamma_1/\gamma_2$. For the pumping coefficients $h_{1,2}$ we employ the normalized pump strength parameters $\eta_1 = 2\mu h_1/\gamma_1\sqrt{\gamma_1\gamma_2}$ and $\eta_2 = 2\mu h_2/\gamma_1\gamma_2$. Lastly, we introduce the normalized increment $y = {\rm i} + \nu/\gamma_1$. The instability occurs for $y'' = {\rm Im}(y) > 1$, so that the threshold equation is $y'' = 1$. 

The phase $\Phi$ is given now by $\Phi = \arg[(\delta_1 - {\rm i})(\delta_2 - {\rm i})]$. Simple relations relevant to the dual backgrounds,
\begin{eqnarray}\label{I1,2}
I_2 &=& I_1^2/4(1 + \delta_2^2) \hspace*{10mm} {\rm FH \, pumping} \\ 
I_2 &=& 1 + \delta_1^2 \hspace*{20mm}  {\rm SH \, pumping} \,,  \nonumber
\end{eqnarray}
enable us to get rid of $I_2$ when analyzing the instability. For the single background in the SH pumping case we have $I^{\rm s}_1 = 0$ and $I^{\rm s}_2 = \eta^2_2/(1 + \delta_2^2)$.

The following relations, replacing Eqs.~(\ref{BackgroundCubic}) and (\ref{F_0^2}), link $I_1$ with $\delta_{1,2}$ and $\eta$ for the FH and SH pumping schemes, respectively:
\begin{eqnarray}\label{NormalizedI}
\eta^2_1 \hspace*{-0.8mm} &=& \hspace*{-0.8mm} \frac{I_1\,[(1 - \delta_1\delta_2 + I_1/2)^2  + (\delta_1 + \delta_2)^2]}{1 + \delta_2^2} \nonumber \\
I^{\pm}_1 \hspace*{-0.8mm} &=& \hspace*{-0.8mm} 2\Big[\delta_1\delta_2 - 1 \pm \sqrt{\eta_2^2 - (\delta_1 + \delta_2)^2} \, \Big].
\end{eqnarray}
In the FH pumping case, the function $I_1(\eta_1)$, single- or three-valued, can be easily quantified for any combination of $\delta_1$ and $\delta_2$; it is not symmetric in $\delta_1$ and $\delta_2$. In the SH pumping case, the lower branch $I_1^-(\eta_2)$ exists only for $\delta_1\delta_2 > 1$; when $\eta_2$ increases from
$|\delta_1 + \delta_2|$ to $\sqrt{(1 + \delta_1^2) (1 + \delta_2^2)}$, $I_1^-(\eta_2)$ changes from $2(\delta_1\delta_2 - 1)$ to $0$. The upper branch $I_1^+(\eta_2)$ exists for any $\delta_{1,2}$. If $\delta_1\delta_2 < 1$, $I_1^+(\eta_2)$ changes from $0$ to $\infty$ when $\eta_2$ grows starting from $\sqrt{(1 + \delta_1^2) (1 + \delta_2^2)}$. If $\delta_1\delta_2 > 1$, $I_1^+(\eta_2)$ changes from $2(\delta_1\delta_2 - 1)$ to $\infty$ when $\eta_2$ grows starting from $|\delta_1 + \delta_2|$. These observations supplement the data of Sect.~\ref{S3}.

\section{Internal instability ($\bm{j = 0}$)}\label{S5}

The case $j = 0$ means that we consider temporal instability of the dual background $\bar{F},\bar{S}$ against
spatially uniform perturbations. Neither dispersion nor walk-off influence this case making it quite general. Here we have for $\gamma_2 = \gamma_1$: $L^{\pm}_{1,2}/\gamma_1 = \pm y - \delta_{1,2}$ leading to a biquadratic equation for $y$ and its solution in the form
\begin{equation}\label{y-FH}
y = \pm \big(p \pm \sqrt{p^2 - 4 q}\;\big)^{\hspace*{-0.3mm} 1/2}/\sqrt{2} \;,
\end{equation}
where $p = \delta_1^2 + \delta_2^2 + 2I_1 - I_2$ and $q = (\delta_1\delta_2 - I_1)^2 -
\delta_2^2I_2$; the signs $\pm$ have to be applied independently. Intensity $I_2$ can be easily excluded from here using Eqs.~(\ref{I1,2}), and intensity $I_1$ can be expressed by $\eta_1$ (or $\eta_2$) and $\delta_{1,2}$ using Eqs.~(\ref{NormalizedI}). 

In the FH pumping case, the lowest threshold value of $\eta_1$ is $[\eta_1^{\rm th}]_{\min} = 6$, it corresponds to $\delta_{1,2} = 0$ and $I_{1,2} = 4$. With increasing $|\delta_{1,2}|$, the threshold value $\eta^{\rm th}_1$ grows rapidly. 
\begin{figure}[h]
	\centering
	\includegraphics[height=4.45cm]{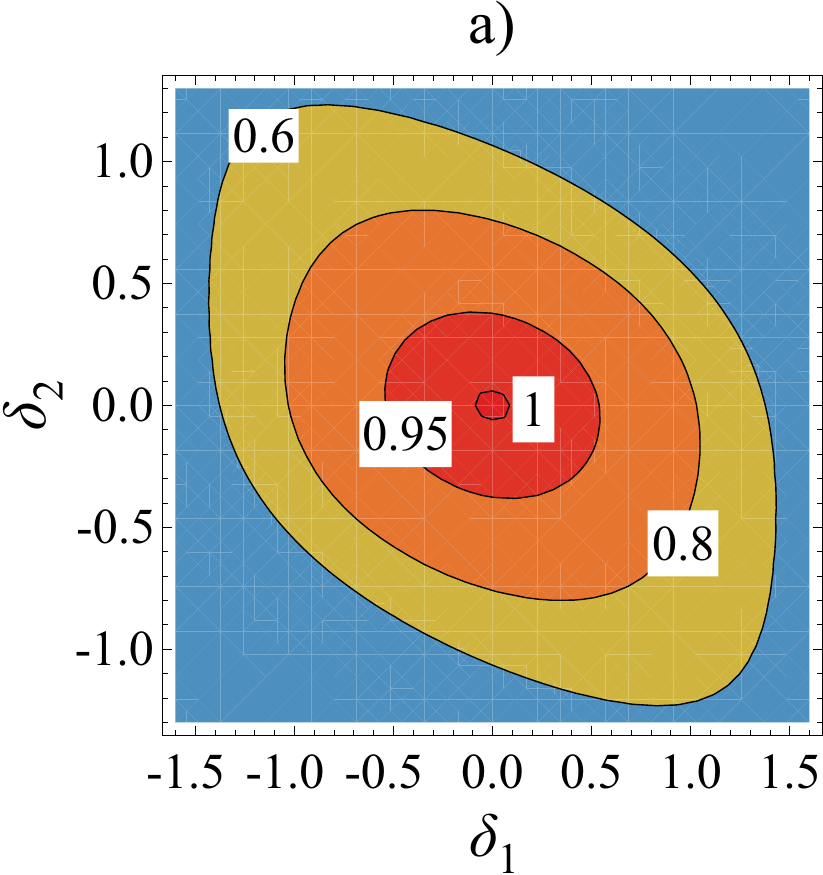} \hspace*{-0.8mm}
	\includegraphics[height=4.45cm]{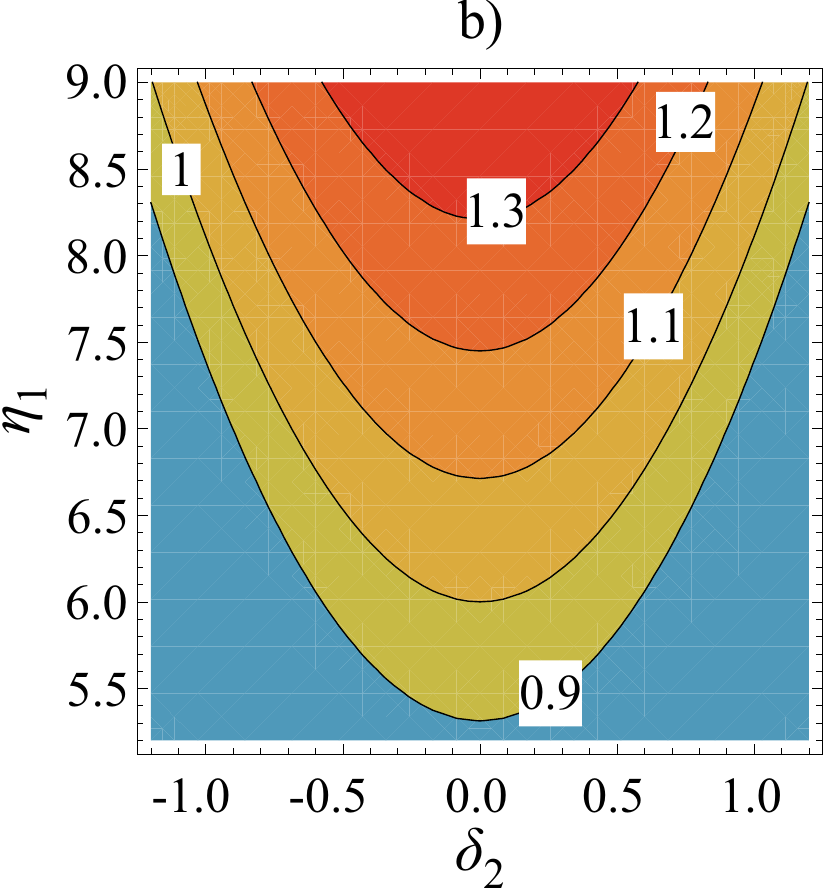}
	\caption{Internal instability for FH pumping. a) Contour lines $y''(\delta_1,\delta_2) = {const}$ for the normalized pumping parameter $\eta_1 = 6.01$ slightly exceeding $[\eta_1^{\rm th}]_{\min} = 6$. b) Contour lines $y''(\delta_2,\eta_1) = {const}$ for $\delta_1 = 0$.
	}\label{FH-Int}
\end{figure}
This is illustrated by Fig.~\ref{FH-Int}a representing contour lines $y''(\delta_1,\delta_2) = {\rm const}$ for $\eta_1 = 6.01$ slightly exceeding the minimum threshold value. One can see only a single contour line with ${\rm const} = 1$, all other lines correspond to a stable background. These data are supplemented by Fig.~\ref{FH-Int}b representing lines $y''(\delta_2,\eta_1) = {const}$ for $\delta_1 = 0$. One sees that the increment grows monotonously with $\eta_1$ and decreases with $|\delta_2|$. Contour lines $y''(\delta_1,\eta_1) = {const}$ for $\delta_2 = 0$ look similarly. The instability occurs equivalently for the $++$ and $--$ combinations of signs in Eq.~(\ref{y-FH}). The real part of the increment, $y'$, is generally nonzero.

A consequence of the above internal instability is the emergence of periodic self-oscillations of zero harmonics $F_0$ and $S_0$ (and of their intensities) above the threshold, $\eta_1 > 6$. These oscillations persist for $r = \gamma_1/\gamma_2 \neq 1$. Also, they do not suffer from the inclusion of nonzero FH and SH harmonics into our considerations. Investigation of the corresponding nonlinear auto-oscillations is, however, beyond the scope of this paper. 

In the SH pumping case, the situation is different. Here the $+-$ combination of signs in Eq.~(\ref{y-FH}) is actual. The whole lower branch $I_1^-(\eta_2)$ is unstable, while the whole upper branch $I_1^+(\eta_2)$ is internally stable. This feature is expected; it leads to the conventional hysteresis behavior when adiabatically increasing and decreasing $\eta_2$.  

All further cases are relevant to the instability against spatio-temporal perturbations -- the external instability.     

\section{The OPO case for $\bm{0,\bar{S}}$ background}\label{S6}

Now we consider the external instability of the single background state $0,\bar{S}$ relevant to the SH pumping scheme, when $h_1 = 0$, $\bar{F} = 0$, and $\bar{S} = -{\rm i}h_2/\Omega_2$. This case corresponds to the conventional OPO. Using Eqs.~(\ref{DispersionEquation}), (\ref{L1,2}) and our normalization, we obtain readily
\begin{equation}\label{OPO1} 
y = \pm {\rm i} \sqrt{I^{\rm s}_2 - (\beta_1j^2 + \delta_1)^2} \;, 
\end{equation}
where $I^{\rm s}_2 = \eta^2_2/(1 + \delta_2^2)$ and $j = 0,\pm 1, \ldots$. It is valid for arbitrary ratio $r = \gamma_1/\gamma_2$ and includes neither the walk-off parameter $\alpha$ nor the dispersion parameter $\beta_2$. The instability occurs for $\eta_2 > \eta_{2,j}^{\rm th} = \sqrt{[1 + (\beta_1j^2 + \delta_1)^2](1 + \delta_2^2)}$ and is relevant to the sign "plus" in Eq.~(\ref{OPO1})~\cite{FootNote2}. Considered as a function of $\delta_1$, $\eta_{2,j}^{\rm th}$ acquires the same minimal value of $(1 + \delta_2^2)^{1/2}$ at $\delta_1 = \delta_{1,j} = -\beta_1 j^2$. Thus, degeneration in $|j|$ takes place. The absolute minimum of $\eta_2^{\rm th}$, $[\eta_2^{\rm th}]_{\min} = 1$, corresponds to $\delta_2 = 0$. For $\beta_1 = d_1/\gamma_1R^2 > 0$, which is relevant to $\lambda_2 \gtrsim 1\,\mu$m in LN based resonators (see also Fig.~\ref{vq}b), the values of $\delta_{1,j}$ are negative. As $\beta_1 \ll 1$, several modes can be excited simultaneously for $\eta_2$ slightly exceeding $1$. For $\delta_1 \geq 0$, the lowest threshold corresponds to $j = 0$; increasing $\eta_2$ is expected to lead here to the excitation of the dual background. 

Within the linear approximation, the quantities $F_j$ and $F^*_{-j}$ are coupled with each other. It is not difficult to obtain that $|F_j| = |F_{-j}|$ above the threshold, and the sum of the phases $\arg(F_j) + \arg(F_{-j})$ is fixed. The phase $\arg(F_j)$ can be treated as a free parameter. 

An important generalization of Eq.~(\ref{OPO1}) has to be mentioned. While the SH amplitude $S(\varphi)$ is always $2\pi$-periodic, the FH amplitude $F(\varphi)$ can be not only periodic, but also antiperiodic~\cite{WePRA20,WeOE20}. This is consistent with periodicity of true light fields and corresponds to pumping into SH modes with even and odd azimuth numbers. The periodic and antiperiodic solutions are topologically different. As $\bar{F} = 0$, we are free to use not only periodic but also antiperiondic FH perturbations.
Generalization of Eq.~(\ref{OPO1}) on the antiperiodic case means that the numbers $j$ are semi-integer, $j = \pm 1/2,\pm 3/2, \ldots$. The threshold properties for the excitation of the periodic and antiperiodic solutions are similar.   

\section{Instability at large walk-off}\label{S7}

We turn next to the external instability of the dual background $\bar{F},\bar{S}$ for large temporal walk-off parameter, $|\alpha| = |v_{12}|/\gamma_2 R \gg 1$. This case is typical for experiments with birefingent LN phase matching at $\lambda_p = \lambda_1 \simeq 1064$~nm~\cite{NaturalPM,IngoPRL20,IngoAPL20}. Here, the velocity difference $v_{12}$ is close to $10^9$~cm/s, both dispersion coefficients are negative, $d_1 \simeq -0.35$ and $d_2 \simeq -0.7$ of $10^4$~cm$^2$/s, and $Q_1 \approx 10\,Q_2$~\cite{IngoAPL20,LN-Methods}. Setting $R = 1$~mm and $\gamma_1 = 10^7$~s$^{-1}$ ($Q_1 \simeq 10^8$) and $\gamma_2 = 2 \times 10^8$~s$^{-1}$ ($Q_2 \simeq 10^7$), we obtain representatively that $\alpha \simeq 50$, $\beta_1 \simeq -0.035$, and $\beta_2 \simeq -0.0035$. Other admittable choices of $R$ and $\gamma_{1,2}$ are able to change notably these estimates. Using quasi-phase matching via the radial poling~\cite{IngoReview16,RadialPoling1,RadialPoling2}, larger and smaller values $|\alpha|$ can be realized for coupling of modes of the same polarization~\cite{WeOE20,WeOE21}. 

Within the simplest approximation we keep only the walk-off contribution to $L_2^{\pm}$ in Eq.~(\ref{L1,2}), thus setting $L^{\pm}_2 = \pm v_{12}j/R$. After that we get from Eqs.~(\ref{DispersionEquation}) for $j \neq 0$: 
\begin{equation}\label{Large0}
y = \frac{I_1}{\alpha j} \pm {\rm i}\sqrt{I_2 - (\beta_1j^2 + \delta_1)^2} \,.
\end{equation}  
The corresponding relation for $y''$ differs from Eq.~(\ref{OPO1}) only by the replacement of the single-background intensity $I_2^{\rm s}$ by the double-background intensity $I_2$. The instability condition $y'' > 1$ is not affected here by~$\alpha$. 

The first term in Eq.~(\ref{Large0}) gives in fact only the first correction to $y$ in $1/\alpha$. The next correction can also be important. To get it, we employ a more accurate approximation for $L_2^{\pm} \simeq $ in Eqs.~(\ref{L1,2}). In the normalized variables, it reads $L^{\pm}_2/\gamma_2 = \pm (\alpha j + {\rm i}) $. To modify Eq.~(\ref{Large0}), it is sufficient to replace there $\alpha j$ by $\alpha j + {\rm i}$. Taking into account that $|\alpha|j \gg 1$ we come to a modified threshold relation
\begin{equation}\label{ThresholdNew}
I_2 = 1 + \big(\beta_1 j^2 + \delta_1 \big)^2 + \frac{2I_1}{\alpha^2j^2} \;.
\end{equation}
It is valid for FH and SH pumping cases regardless of the ratio $r = \gamma_1/\gamma_2$. Validity of this threshold relation and its high accuracy have been verified by direct solving of Eqs.~(\ref{DispersionEquation}) and (\ref{NormalizedI}). The last term in Eq.~(\ref{ThresholdNew}), i.e. the walk-off correction, tends to increase the threshold. 

\begin{figure}[h]
\centering
\includegraphics[width=4.47cm]{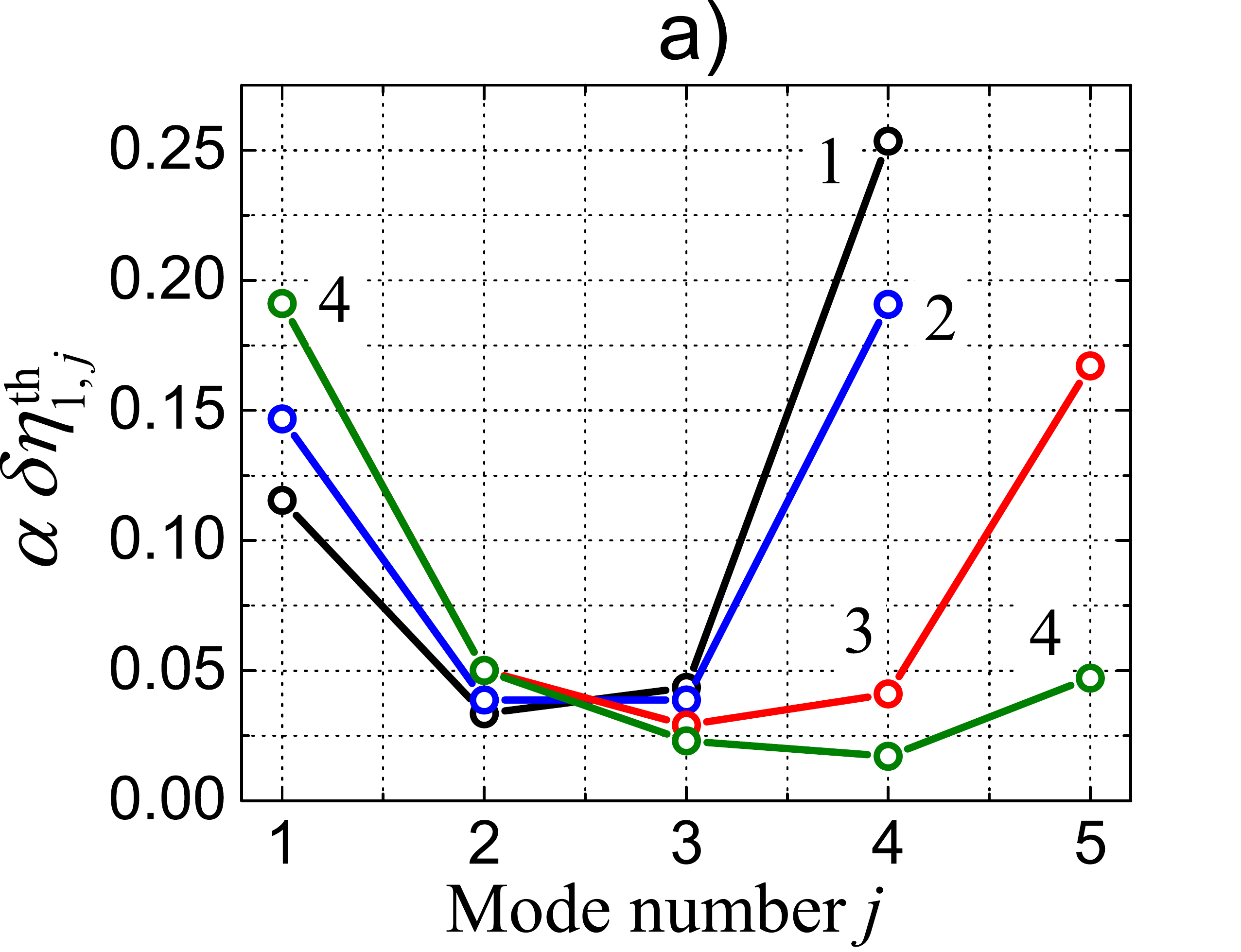} \hspace*{-5mm}
\includegraphics[width=4.47cm]{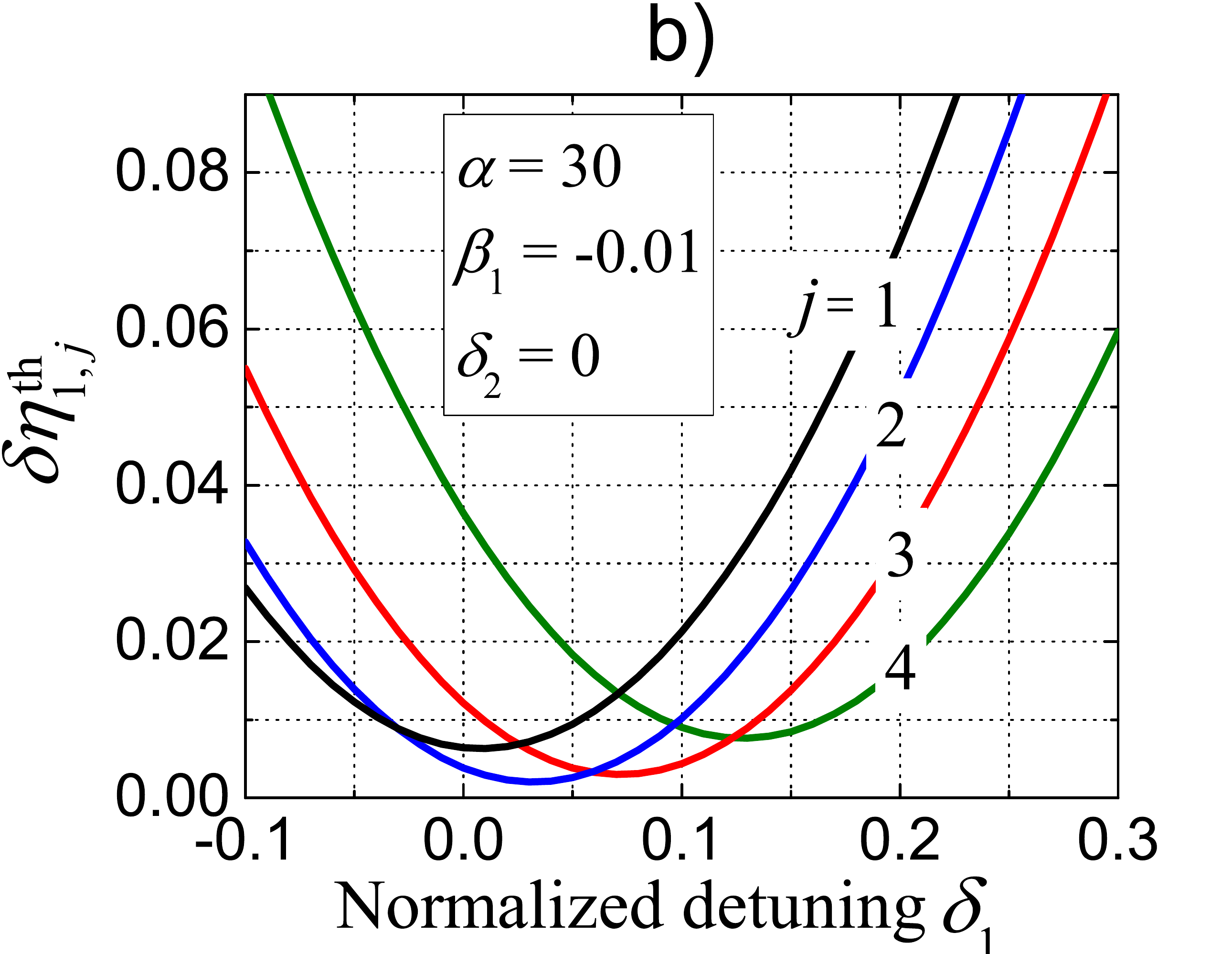}
\caption{a) Minimal values of $\delta \eta_{1,j}^{\rm th}(\delta_1,\delta_2)$ for several first mode numbers and different pairs $(|\alpha|,|\beta_1|)$: Circles $1$, $2$, $3$, and $4$ correspond to the pairs $(50,0.03)$, $(40,0.03)$, $(30,0.03)$, and $(30,0.01)$. The lines serve as a guide to the eye. b) Representative dependences of $\delta \eta_{1,j}^{\rm th}(\delta_1,0)$ for $\alpha = 30$, $\beta_1 = -0.01$, and $j = 1,2,3$, and $4$. The lowest threshold corresponds to $|j| = 2$. 
	}\label{LargeAlpha}
\end{figure}

Turning to the consequences of Eq.~(\ref{ThresholdNew}), we consider first the FH pumping case. As $I_2 = I_1^2/4(1 + \delta_2^2)$ and $\eta_1 = \eta_1(I_1,\delta_1,\delta_2)$ according to Eqs.~(\ref{I1,2}) and (\ref{NormalizedI}), we have an implicit relation for $\eta_1^{\rm th}$ as a function of $\delta_{1,2}$ for each $j$. Taking now into account that $|\alpha| \gg 1$ and $|\beta_1| \ll 1$, we see that the last two terms in Eq.~(\ref{ThresholdNew}) are relatively small for $|\delta_{1,2}| \ll 1$ and not too large $|j|$. Within this important range of parameters, we use a simple perturbative approach. One can check first using Eqs.~(\ref{NormalizedI}) and~(\ref{ThresholdNew}) that at the threshold $\eta_1 = 2\sqrt{2}$, $I_1 = 2$, and $I_2 = 1$ in the leading approximation. Setting next $\eta_1 = 2\sqrt{2} + \delta\eta_1$ and $I_1 = 2 + \delta I_1$, we obtain that $\delta I_1 = \delta\eta_1/\sqrt{2} + \delta_2^2 - (\delta_1 - \delta_2)^2/4$ and 
\begin{equation}\label{FH-Correction}
\frac{\delta \eta_{1,j}^{\rm th}}{\sqrt{2}} = (\beta_1 j^2 + \delta_1)^2 + \frac{(\delta_1 - \delta_2)^2}{4} + \frac{4}{\alpha^2j^2} \;. 
\end{equation}
The function $\delta \eta_{1,j}^{\rm th}(\delta_1,\delta_2)$ has a minimum at $\delta_{1,2} = |\beta_1|j^2$ with the minimal value $4\sqrt{2}/\alpha^2j^2$. The larger $|j|$, the lower is this value. However, for $|j| \gtrsim 1/\sqrt{|\beta_1|}$, i.e. for $|j|> 4$-$5$, our approximation breaks and we should substitude the exact dependence $I_1(\eta_1,\delta_{1,2})$, given by Eq.~(\ref{NormalizedI}), into Eq.~(\ref{ThresholdNew}). As the result, the minimal value of $\eta_{1,j}^{\rm th}(\delta_1,\delta_2)$ starts to grow for sufficiently large $|j|$, and this effect depends on $\alpha$ and $\beta_1$, as illustrated by Fig.~\ref{LargeAlpha}a.
One sees that, depending on $\alpha$ and $\beta_1$, the lowest threshold corresponds to different modal numbers. This correlates with the experimental data of~\cite{IngoAPL20}. For realistic parameters, the optimum value of $j$ cannot, however, be very large. Another playable variant of detuning adjustment is $\delta_2 = 0,\delta_1 \neq 0$. In particular, the case of perfect phase matching, where $\Delta_2/\Delta_1 = 2$, corresponds to $\delta_2/\delta_1 = Q_2/Q_1 \ll 1$. According to Eq.~(\ref{FH-Correction}), the function $\eta_{1,j}^{\rm th}(\delta_1,0)$ has a minimum at $\delta_1 = 0.8|\beta_1|j^2$ with the minimal value $(\sqrt{2}\beta_1^2/5)\,(j^2 + 20/\alpha^2\beta_1^2j^2)$. The $j$-dependence of this value is controlled by the product $|\alpha\beta_1|$. The smaller this product, the larger is the optimum $j^2$. This also makes us to use the exact relation~(\ref{NormalizedI}) for $I_1(\eta,\delta_1)$ instead of Eq.~(\ref{FH-Correction}) to optimize $j^2$. Figure~\ref{LargeAlpha} illustrates the results of this consideration. For the value $|\alpha \beta_1| = 0.3$ representing an optimistic estimate, the optimum value of $j$ is $2$. 

Consider now the SH pumping case. Neglecting first the correction $\delta y \propto 1/\alpha^2$ and substituting Eq.~(\ref{I1,2}) for $I_1$ into (\ref{Large0}), we obtain that $y'' = \pm \sqrt{1 - \beta_1^2j^4 - 2\beta_1\delta_1 j^2}$. The instability occurs for opposite signs of $\beta_1$ and $\delta_1$ and is restricted to the mode numbers $|j| < |2\delta_1/\beta_1|^{1/2}$. For $\delta_1/|\beta_1| > 1/2$, the dual background is stable. 
Remarkably, the pump strength parameter $\eta_2$ does not enter the above relation for $y''$. However, one should keep in mind that $\eta_2$ is restricted from below by the inequalities indicated in the end of Sect.~\ref{S4}. In particular, for $\delta_1\delta_2 < 1$ and $\eta_2^2 \to (1 + \delta_1^2)(1 + \delta_1^2)$, when $I^+_1 \to 0$, we have $I_2 = I^{\rm s}_2$. Thus, we have a continuous transition to the OPO case, see Sect.~\ref{S6}. In fact, this transition is valid for any $\alpha$ in the SH case. This is evident from Eq.~(\ref{DispersionEquation}): For $|\bar{F}| \to 0$, parameters $L_2^{\pm}$ (relevant to the walk-off) cancel out.  

The correction $\delta y''$ relevant to the walk-off parameter leads merely to a slight narrowing of the above considered instability region. It does not influence the limit $I_1^+ \to 0$.

\section{Zero walk-off case}\label{S8}

The case $\alpha = v_{12}/\gamma_2 R = 0$ ensures the maximum involvement of the SH perturbations with $j = \pm 1, \pm 2, \ldots$ in the interaction. On the other hand, it is promising for realization of $\chi^{(2)}$ soliton-combs states~\cite{WeOE20,WeOE21}. Here, the main question is about the effect of this choice on the external instability of the dual background. At $\alpha = 0$ and $\gamma_1 = \gamma_2$, we also arrive from Eqs.~(\ref{DispersionEquation}),~(\ref{L1,2}) to a biquadratic equation for the increment $y$. Its solution is given by Eq.~(\ref{y-FH}) with parameters
\begin{eqnarray}\label{p0,q0}
\hspace*{-3mm} p \hspace*{-1mm} &=& \hspace*{-1mm} (\delta_1 + \beta_1j^2)^2
\hspace*{-0.5mm} + \hspace*{-0.5mm}
(\delta_2 + \beta_2j^2)^2 \hspace*{-0.5mm} + \hspace*{-0.5mm} 2I_1 \hspace*{-0.5mm} - \hspace*{-0.5mm} I_2 \\
\hspace*{-3mm} q \hspace*{-1mm} &=& \hspace*{-1mm} [(\delta_1 + \beta_1j^2)(\delta_2 +
\beta_2j^2) \hspace*{-0.5mm} - \hspace*{-0.5mm} I_1]^2 \hspace*{-0.5mm} - \hspace*{-0.5mm}
I_2 (\delta_2 \hspace*{-0.5mm} + \hspace*{-0.5mm} \beta_2j^2)^2 \nonumber
\end{eqnarray}
and $j = \pm 1, \pm 2, \ldots$. Setting formally $j = 0$ we return to the case of internal instability, Sect.~\ref{S5}. Intensity $I_2$ is expressible by $I_1$ with Eqs.~(\ref{I1,2}), and $I_1$ can be expressed by $\eta_1$ (or $\eta_2$) and $\delta_{1,2}$ using Eqs.~(\ref{NormalizedI}). 

For LN crystals we have $d_1 \simeq 0.4 \times 10^4$ and $d_2 \simeq -0.2 \times 10^4$~cm$^2$/s at the zero walk-off point $\lambda_2 \equiv \lambda_1/2 = \lambda_c \simeq 1349$~nm, see also Fig.~\ref{vq}. In contrast to the case of birefringent phase matching (Sect.~\ref{S7}), coefficient $\beta_1$ is positive. As the normalized dispersion coefficients we set representatively $\beta_1 = -2\beta_2 = 0.06$. This corresponds, in particular, to $R = 1$~mm and $Q_2 = 2Q_1 \simeq 10^8$. 

\begin{figure}[h]
	\centering \hspace*{-1mm}
	\includegraphics[height=4.7cm]{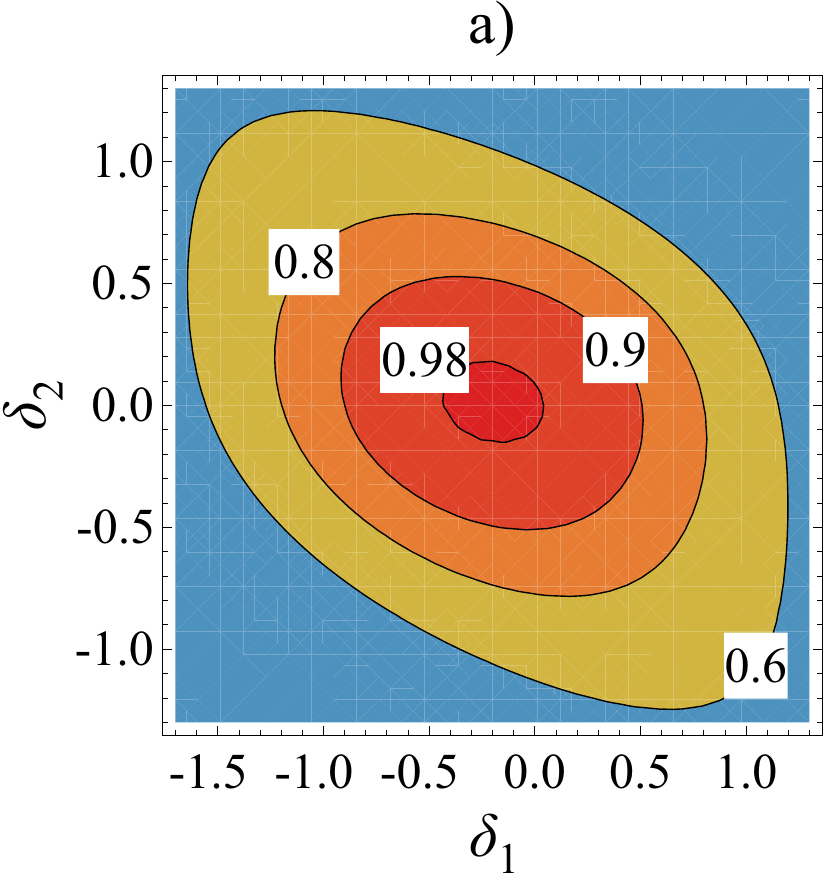}
	\includegraphics[height=4.7cm]{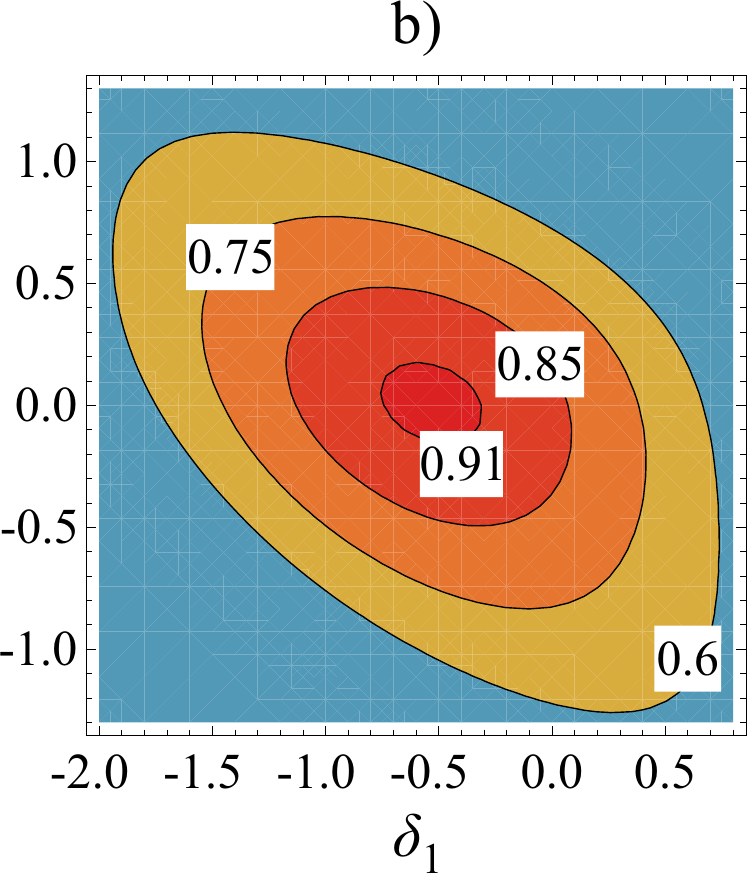}
	\caption{The FH pumping case at $\alpha = 0$: Contour lines $y''(\delta_1,\delta_2) = {\rm const}$ for $\eta_1 = 6.01$ exceeding by $0.01$ the internal instability threshold and two values of the modal number, $j = 3$ (a) and $5$ (b). Note different horizontal scales. The values of $y''_{\max}$ in a) and b) are $\simeq 0.989$ and $0.917$, respectively. 
	}\label{FHj}
\end{figure}

In the FH pumping case we have found that all modes with $j \neq 0$ possess higher instability thresholds than the mode with $j = 0$. This is established by plotting contour lines $y''(\delta_1,\delta_2) = {\rm const}$ for different values of $\eta_1$ and $j$ and illustrated by Fig~\ref{FHj} for $j = 3$ and $5$. 
Here we set, as in Fig.~\ref{FH-Int}a, $\eta_1 = 6.01$ exceeding by $0.01$ the minimum threshold value for the internal instability. However, in contrast to Fig.~\ref{FH-Int}a, the contour line with ${\rm const} = 1$ corresponding to the instability threshold is absent. All contour lines are shifted to the left about the vertical $\delta_1 = 0$ and the shift grows with $j$. This is similar to the OPO case with $\beta_1 > 0$. However, the values of $|\delta_1|$ relevant to the maxima in a) and b), $\simeq 0.2$ and $0.55$, are substantially larger than the OPO related values $\beta_1j^2$. Detunings $\delta_{1,2}$ relevant to Fig.~\ref{FHj} cannot be treated perturbatively as small quantities. The fact that the lowest instability threshold corresponds to $j = 0$ indicates that development of spatially uniform auto-oscillations of $|F_0|$ and $|S_0|$ is likely for $\alpha \simeq 0$ and $\eta_1 > 6$. 

The fact that the instability thresholds in the FH pumping case are substantially higher for $\alpha = 0$ than they are for $|\alpha| \gg 1$ may look surprising. It means merely that strong involvement of side SH harmonics $S_l$ with $l \neq 0$ results in a destructive interference between nonlinear sum and difference frequency generation processes.

\begin{figure}[h]
	\centering
	\includegraphics[height=3.95cm]{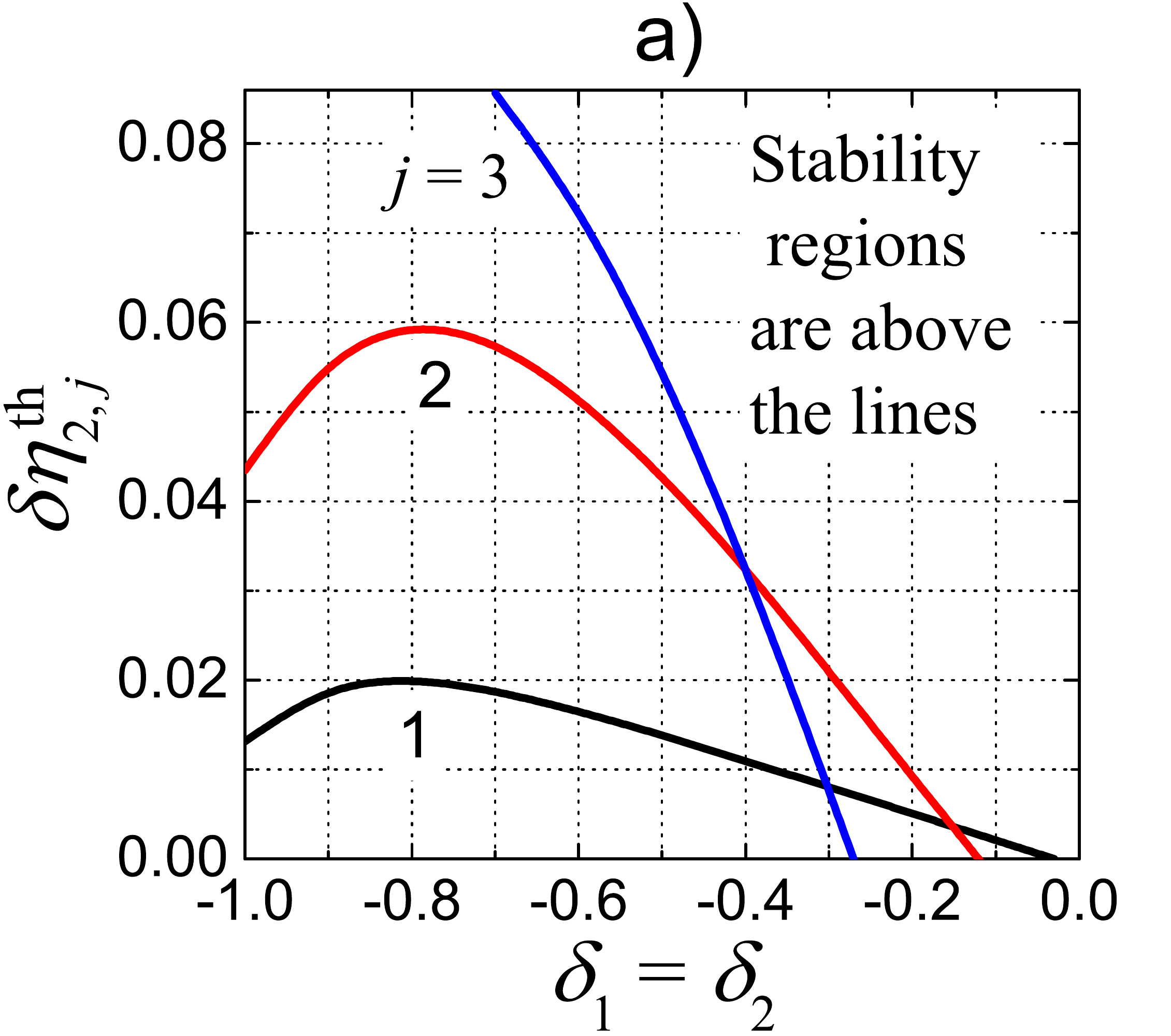} \hspace*{-4mm}
	\includegraphics[height=3.95cm]{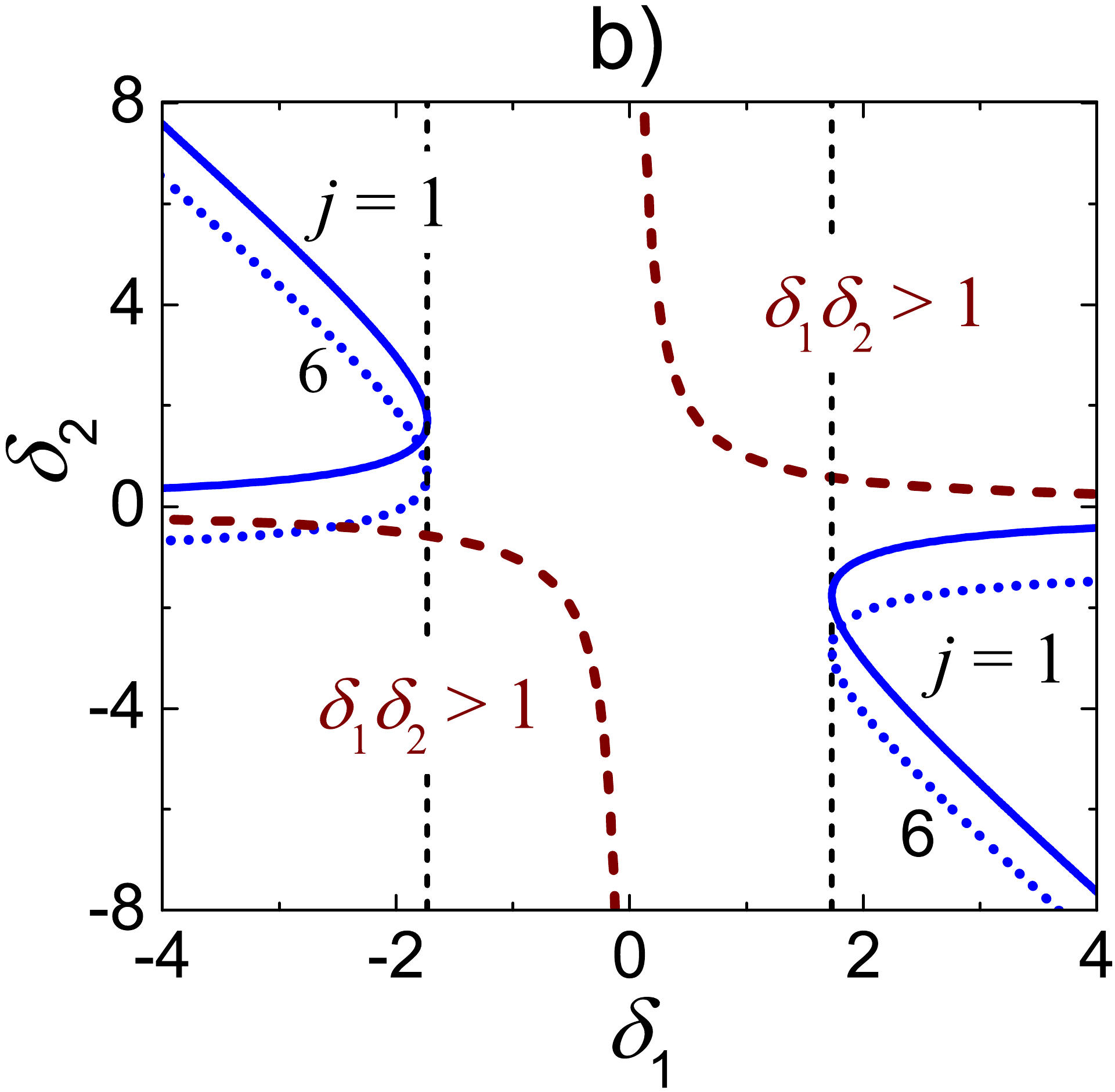}
		\caption{SH pumping: (a) Threshold dependence $\delta\eta_{2,j}^{\rm th}(\delta_1)$ for $\delta_2 = \delta_1$ and $|j| = 1$, $2$, and $3$ in the region of small $I^+_1(\eta_2)$. The regions of stability are above the threshold lines. (b) Regions of stability and instability of the dual background on the $\delta_1,\delta_2$ plane for large values of $I_1(\eta_2)$; solid and dotted blue lines correspond to $|j| = 1$ and $6$. The dashed burgundy lines show the border between the regions with $\delta_1 \delta_2 \lessgtr 1$.  
	}\label{Alpha0}
\end{figure}

In the SH pumping case, the situation is strongly different. First of all, we have to exclude the lower dual background branch $I_1^-$ from our considerations because of its internal instability. For the upper internally stable branch, the dependence $I_1(\eta_2,\delta_1,\delta_2)$ has been considered in some detail in Sect.~\ref{S4}. In the limit $I^+_1 \to 0$, i.e. for $\delta_1\delta_2 < 1$ and $\eta_2 \to \eta_2^{\min} = \sqrt{(1 + \delta_1^2)(1 + \delta_2^2)}$, we return to the case considered in Sect.~\ref{S7}. The corresponding limiting solution for $y''$ follows from Eqs.~(\ref{y-FH}),~(\ref{p0,q0}) for the $+-$ combination of the signs: $y''_{+-} = \sqrt{1 - 2\delta_1\beta_1j^2 - \beta_1^2j^4}$. As already mentioned, it is consistent with Eq.~(\ref{OPO1}) for the OPO case. The instability condition $y''_{+-} > 1$ is fulfilled for sufficiently large negative detuning $\delta_1$. The threshold dependences of $\delta \eta_{2} = \eta_2 - \eta_2^{\min}$ calculated for $\delta_2 = \delta_1$, $j = 1,2,3$, and finite values of $I^+_1$ are presented in Fig.~\ref{Alpha0}a. The points $\delta \eta_{2,j}^{\rm th}(\delta_1) = 0$ correspond to the limit $I_1 \to 0$. Remarkably, increase of $\delta \eta_2$ quickly stabilizes the instability. The threshold curves for $\delta_2 = 0$ look similar.

The limit $I_1 \to 0$ cannot be applied in the region $\delta_1\delta_2 > 1$ where the dependence $I_1(\eta_2)$ starts from a nonzero value $2(\delta_1\delta_2 - 1)$. Furthermore, finite and large values of $I_1(\eta_2)$ are relevant also to the case $\delta_1\delta_2 < 1$, $\eta_2 \gg |\delta_1 + \delta_2|$. For large $I_1(\eta_2)$, another limiting case in Eq.~(\ref{y-FH}) corresponding to the $++$ combination of signs, is important. One sees that the $I_1^2$ terms in the expression for $p^2 - 4q$ cancel out, so that the term proportional to $I_1$ becomes leading inside the radical expression. With the use of Eqs.~(\ref{p0,q0}), this leads ultimately to a simple $I_1$-independent asymptotic expression for the increment: $2y''_{++} = \sqrt{1 + \delta_1^2 - (\delta_1 + \delta_2 + \beta j^2)^2}$ with $\beta = \beta_1 + \beta_2$. The condition $y''_{++} > 1$ determines relatively narrow areas on the $\delta_1,\delta_2$ plane where the dual background with sufficiently large values of $I_1$ is unstable. The borders of these areas for $|j| = 1$ are presented by blue solid lines in Fig.~\ref{Alpha0}b. The extreme left and right points of the instability regions correspond to $\delta_1 = \pm\sqrt{3}$. 
Increase of $|j|$ results in small parallel down shifts of the borders of this region, as exemplified by dotted blue lines for $|j| = 6$. For intermediate values of $I_1$ the instability regions of Fig.~\ref{Alpha0}b shrink. Thus, within broad areas of $\delta_{1,2}$ and $\eta_2$ the dual background remains stable. Fine details of the dependence $y''(\eta_2,|j|,\delta_1,\delta_2)$ are complicated, they are beyond this study. 

\section{Intermediate walk-off range}\label{S9}

To investigate the intermediate range of $\alpha$, we solved Eq.~(\ref{DispersionEquation}) numerically in the FH pumping case. Generally, it is necessary to take into account that the velocity difference $v_{12}$ and the dispersion parameters $d_{1,2}$ are functions of $\lambda_2 \equiv \lambda_1/2$, so that variations of $\alpha(\lambda_2)$ and $\beta_{1,2}(\lambda_2)$ are correlated. However, in the actual case of the extraordinary polarization of the FH and SH modes, the velocity difference $v_{12}(\lambda_2)$ changes very rapidly in the vicinity of zero walk-off point $\lambda_c \simeq 1349$~nm, $v_{12} \, {\rm [cm/s]} \simeq -6 \times 10^5 (\lambda_2 - \lambda_c) \, {\rm [nm]}$. The changes of $\beta_{1,2}(\lambda_2)$ can be neglected here until $|\alpha|(\lambda_2) \lesssim 10^2$ (see also Fig.~\ref{vq}) and we set, as in Sect.~\ref{S8}, $\beta_1 = -2\beta_2 = 0.06$. As the threshold values are even in $\alpha$, we consider only the range $\alpha > 0$.   

\begin{figure}[h]
	\centering
	\includegraphics[height=4cm]{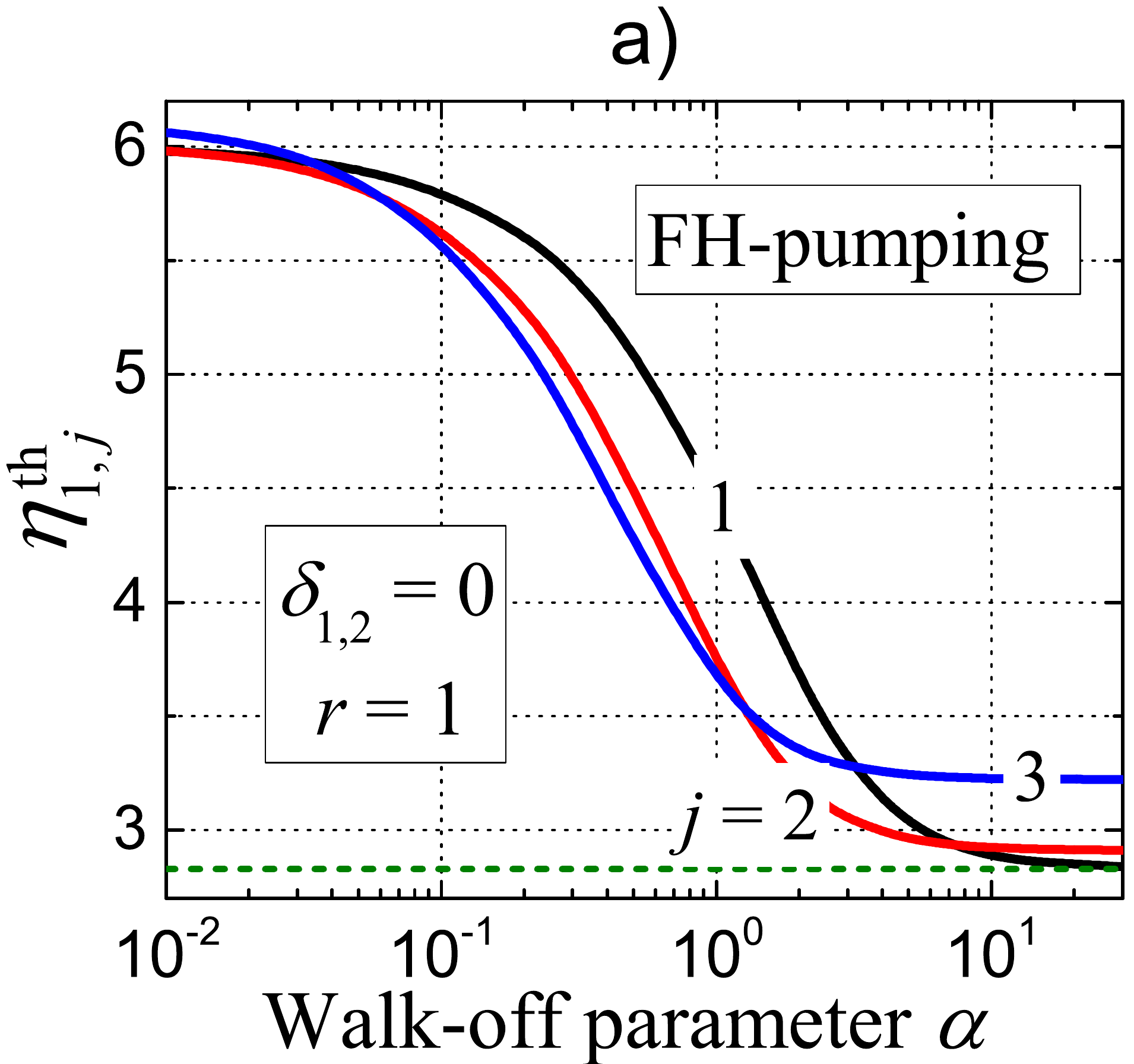} \hspace*{-1mm}
	\includegraphics[height=4cm]{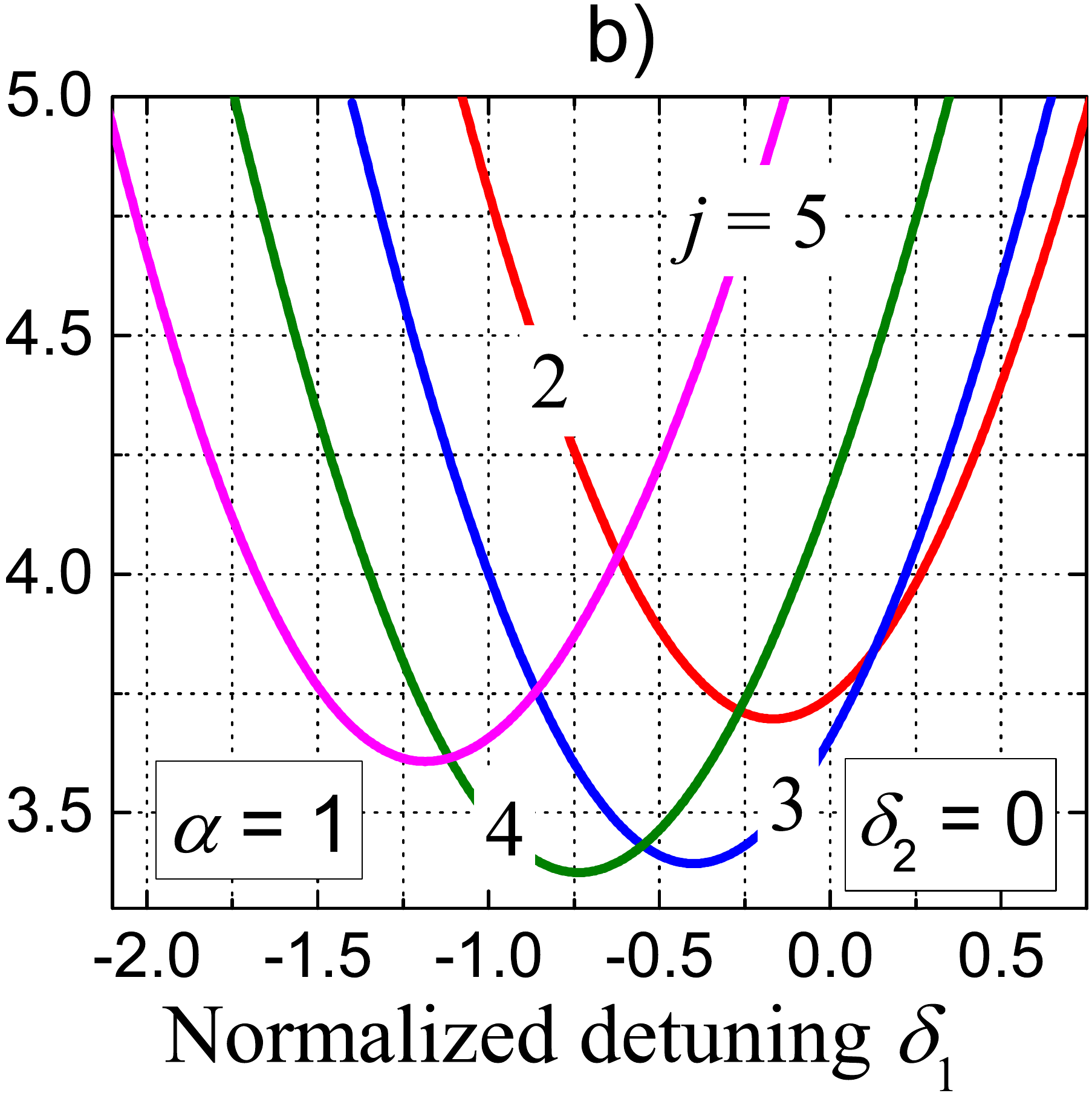}
	\caption{Representative threshold dependences of the normalized pump strength $\eta_1$ in the FH pumping case: a) $\eta^{\rm th}_{1,j}$ versus the walk-off parameter $\alpha$ for $\delta_{1,2} = 0$ and $j = 1,2,3$. b) $\eta^{\rm th}_{1,j}$ versus the normalized detuning $\delta_1$ for $\alpha = 1$, $\delta_2 = 0$, and $j = 2,3,4,5$. The horizontal dashed line corresponds to the level of $2\sqrt{2}$. 
	}\label{Intermediate}
\end{figure}

We analyze first the threshold dependences $\eta^{\rm th}_{1,j}(\alpha)$ relevant to FH pumping at zero detunings. Figure~\ref{Intermediate}a shows them for $j = 1,2,3$ on a logarithmic horizontal scale. 
One sees a continuous transition from the case $\alpha \ll 1$, where $\eta_{1,j}^{\rm th} \simeq 6$, to the case $\alpha \gg 1$, where $\eta_{1,j}^{\rm th} > 2\sqrt{2}$. As we know from Sects.~\ref{S5},~\ref{S8}, the threshold of the internal instability $\eta_{1,0}^{\rm th} = 6$ is the lowest for $\alpha = 0$, i.e. we expect that $\eta^{\rm th}_{1,j}(\alpha)$ with $j \neq 0$ becomes larger than $6$ for $\alpha \to 0$. This expectation is indeed met. However, the thresholds $\eta^{\rm th}_{1,j}(\alpha)$ with $j \neq 0$ become smaller than $6$ already for small values of $\alpha$; the larger $j$, the larger is the necessary value of the walk-off parameter. 
For $j = 1$, this value is $\alpha \approx 10^{-3}$. This feature indicates that auto-oscillations relevant to the internal instability at $\delta_{1,2} = 0$ can exist only in a very close vicinity of the zero walk-off point $\lambda_c$. For modest values of $\alpha$, the value of $j$ minimizing the instability threshold ranges between $3$ and $1$. For $\alpha \gtrsim 10$, it equals $1$, which is consistent with Eq.~(\ref{FH-Correction}).    

Consider now the effect of detuning $\delta_1$ on the threshold values $\eta_{1,j}^{\rm th}$. It is illustrated by Fig.~\ref{Intermediate}b for $\alpha = 1$ and $j = 2,3,4,5$. Negative nonzero detunings substantially and selectively in $j$ lower the threshold power parameter. Such lowering is indeed due to the positive sign of the dispersion coefficient $\beta_1$. Modes with $j = 3$ and $4$ possess here the lowest thresholds. The values $\delta_1^{(j)}$ minimizing the instability thresholds $\eta_{1,j}^{\rm th}(\delta_1)$ can be estimated as $\approx 0.8 \beta_1 j^2$ as in the case $\alpha \gg 1$, see Sect.~\ref{S7}. Almost the same values of $\delta_1^{(j)}$ and $j$ minimizing the thresholds are relevant to $\alpha = 0.1$; the corresponding thresholds, however, are much higher here. 

\section{Discussion}\label{S10}
 
Dual FH-SH background states $\bar{F},\bar{S}$ in monochromatically pumped $\chi^{(2)}$ microresonators obey relations including three variable parameters (the pump strength and two normalized frequency detunings) and involving neither the dispersion nor walk-off parameters. These relations are different in the FH and SH pumping cases. They lead generally to multivalued states. The dual backgrounds are closely related to the $\chi^{(2)}$ comb generation and the SH generation in microresonators.   

There is no doubt that the instability against the excitation of the resonator modes with modal numbers different from those linked primarily by the PM conditions is the starting point of the comb-soliton formation for both pumping schemes. Also, it is an important  factor restricting an efficient SH generation. However, the role and nature of the instability are far from being uniform. They are different for the FH and SH pumping cases and, furthermore, cannot be reduced to the OPO. 

For the FH pumping scheme, the presence of the dual background is inevitable starting from weak pump powers. As often believed, the SH amplitude $\bar{S}$ makes possible an internal OPO above a certain power threshold, while the presence of the FH counterpart $\bar{F}$ is unimportant. However, this is typically not the case. As soon as a weak seed FH perturbation $\delta F$ emerges, it forces a SH perturbation $\delta S$ in a thresholdless manner just because of nonzero $\bar{F}$. As the result, the instability becomes drastically different from the OPO. It involves generally not only the FH dispersion parameters, but also the walk-off parameter and SH dispersion coefficient. Remarkably, the presence of large walk-off, typical of the birefringent (natural) PM, makes the situation similar to the OPO case. On the other hand, large FH-SH velocity difference hinders efficient generation of the dual combs~\cite{WeOE20,WeOE21}.

Vicinity of the point of zero walk-off, which is the most favorable for the $\chi^{(2)}$ comb generation and accessible via the radial poling~\cite{RadialPoling1,RadialPoling2}, is highly special for the dual background instability in the FH pumping case. While exactly at this point ($\alpha = 0$) the background is unstable only internally under rather large pump powers, small deviations ($|\alpha| \ll 1$) lower the power threshold and lead to the excitation of modes with $|j| \neq 0$. The larger the deviation $|\alpha|$, the lower is the threshold and larger is number $j$. Development of such a modulation instability is expected to lead to the formation of different periodic multisoliton states. 

For the SH pumping scheme, the situation is different. Here the single SH background $0,\bar{S}$ exists for sufficiently low pump powers. Owing to the absence of the FH counterpart ($\bar{F} = 0$), SH perturbations are absent within the linear approximation, and the true OPO occurs above a certain threshold. Just this process initiates different nonlinear states. Importantly, not only periodic but also antiperiodic (with semi-integer $j$) FH perturbations are allowed in this case. Within a broad range of variable parameters (the pump power, the PM wavelength, the frequency detunings) the antiperiodic soliton-comb states are predicted to be stable and easily accessible above the OPO threshold~\cite{WeOE20,WeOE21}. The left and right asymptotic values of the FH amplitude $F(\varphi)$ achieved far from the soliton core are just $\pm{\bar{F}}$.

Stability of the dual FH-SH backgrounds is a necessary condition for the existence of both periodic and antiperiodic soliton-comb states. This condition, as we have seen, is fulfilled within broad ranges of the variable parameters in the SH pumping case. There are, however, windows of these parameters where the dual background is unstable and, therefore, comb-soliton generation is forbidden.   

In contrast to numerical studies of nonlinear regimes involving many combinations of the input parameters, our analysis of the linear stability of the dual FH-SH backgrounds is relatively simple, and it greatly relies on analytical methods. The characteristic 4th degree equation for the increment is general for the FH- and SH-pumping cases. It involves all relevant input parameters and  can be investigated with elementary numerical tools. On the other hand, it cannot be reduced to the known relations relevant to the conservative $\chi^{(2)}$ case~\cite{SkryabinReview}. 

\section{Conclusions}\label{S11}

\noindent -- Simple relations, different for the FH and SH pumping, control the dependence of dual backgrounds on the variable parameters of $\chi^{(2)}$ microresonators. They are related to the comb and SH generation. \\
-- Internal and external instability of the dual backgrounds is investigated within the linear approximation in spatio-temporal perturbations.  \\
-- The external instability, involving the excitation of new modes, is strongly different from the OPO. It involves the temporal walk-off between the FH and SH modes. \\
-- Different particular and limiting cases are considered, and the regions of instabilities in the space of variable parameters are determined. Vicinity of the zero walk-off point is found to be special for the FH and SH pumping schemes.


\begin{thebibliography}{99}
\bibitem{FootNote2}
The use of subscript "th" should be commented. Generally, the threshold condition links three independent variable parameters: $\eta_{1,2}$, $\delta_1$, and $\delta_2$. Notation $\eta_{1,2}^{\rm th}$ is used to stress the importance of the pump strength parameter.
\end{thebibliography}
\end{document}